\documentclass[twocolumn,showpacs,floats,floatfix,superscriptaddress,aps,pra]{revtex4}

\usepackage{amsfonts,amssymb,amsmath}
\usepackage{color,calc}
\usepackage[dvips]{graphicx}
\usepackage{bm}
\usepackage{citesort}

\def\be{ \begin{equation} }
\def\ee{ \end{equation} }
\def\bea{ \begin{eqnarray} }
\def\eea{ \end{eqnarray} }
\def\bse{ \begin{subequations} }
\def\ese{ \end{subequations} }

\def\jones{\mathbf{J}}
\def\ret0{\varphi_0}

\def\black{}

\def\ttl#1{``#1''}
\def\ttl#1{}
\def\retardation{\Phi}
\def\eps{\epsilon}
\def\m{m}

\begin{document}

\title{Highly efficient broadband polarization retarders and tunable polarization filters made of composite stacks of ordinary wave plates}

\author{Emiliya St. Dimova}\email{edimova@issp.bas.bg}
\affiliation{Institute of Solid State Physics, BAS, 72 Tzarigradsko Chauss\'ee Blvd, 1784 Sofia}
\author{Svetoslav S. Ivanov}
\affiliation{Department of Physics, St. Kliment Ohridski University of Sofia, 5 James Bourchier Blvd, 1164 Sofia, Bulgaria}
\author{George St. Popkirov}
\affiliation{Central Laboratory for Solar Energy and New Energy Sources, BAS, 72 Tzarigradsko Chauss\'ee Blvd, 1784 Sofia}
\author{Nikolay V. Vitanov}
\affiliation{Department of Physics, St. Kliment Ohridski University of Sofia, 5 James Bourchier Blvd, 1164 Sofia, Bulgaria}

\begin{abstract}
By using the formal analogy between the evolution of the state vector in quantum mechanics and the Jones vector in polarization optics,
we construct and demonstrate experimentally efficient broadband half-wave polarization retarders and tunable narrowband polarization filters.
Both the broadband retarders and the filters are constructed by the same set of stacked standard multi-order optical wave plates rotated at different angles with respect to their fast polarization axes:  for a certain set of angles this device behaves as a broadband polarization retarder while for another set of angles it turns into a narrowband polarization filter.
We demonstrate that the transmission profile of our filter can be centered around any desired wavelength in a certain vicinity of the design wavelength of the wave plates solely by selecting appropriate rotation angles.
\end{abstract}

\maketitle

\section{Introduction}
Optical polarization retarders have countless applications in physical experiments.
Retarders are optical plates, made of birefringent material, which possess two orthogonal axes --- ordinary and extraordinary --- with different refractive indices, $n_{o}$ and $n_{e}$.
Light of wavelength $\lambda$ passed orthogonally through a retarder of thickness $L$ experiences a retardation (phase shift) $\varphi=2\pi L(n_{e}-n_{o})/\lambda$.
Common types of wave plates (WPs) are half-wave plates (HWP) with retardation $\varphi=(\m+1)\pi$  and quarter-wave plates (QWP) with $\varphi=(\m+\tfrac12)\pi$, $(\m=0,1,2,\ldots)$.
The QWPs and HWPs can be zero-order $(\m=0)$ or multi-order $(\m>0)$.
Because $\varphi$ depends on $\lambda$, the retarders exhibit wavelength dispersion and therefore QWP and HWP are
produced with a specified design wavelength.
Applications, e.g. in the terahertz time-domain spectroscopy \cite{THz1, THz2}, microwave
polarimetry \cite{MicroPolar1, MicroPolar2, MicroPolar3}, etc. often demand wider wavelength range of nearly
constant phase retardation.

Achromatic (or broadband, BB) retarders were first proposed by West and Makas \cite{West49} by
combining two or more plates with different birefringence, whereby one of the plates is rotated to a specified angle.
Destriau  and Prouteau \cite{Destriau49} used two WPs of the same material but with different thickness and axis angles.
Pancharatnam used three plates to construct half-wave \cite{Pancharatnam55a} and quarter-wave \cite{Pancharatnam55b} BB retarders.
Harris \emph{et al.} proposed achromatic QWPs with 6 \cite{Harris64} and 10 \cite{McIntyre68} stacked identical zero-order QWPs rotated at different angles, optimized to
reduce the wavelength dispersion over an extended range.
Alternatively, for some applications tunable phase retardation plates can be used.
Polarization retarders can be made tunable either
by tilting the WP by a specified angle (Alphalas GmbH, G\"{o}ttingen)
or electrically, by changing the birefringent optical path length using a twisted nematic liquid crystal phase retarder \cite{LiqCr1, LiqCr2}.

Birefringent phase retarders have been also used to construct narrowband (NB) optical filters.
The Lyot \cite{L44, T01} and $\check{\text{S}}$olc filters \cite{E58, T01, S65} are build up by stacks of retarders and polarizers.
Possibilities to obtain very narrowband transmission filters for applications, e.g. solar imaging at specific wavelengths, were also demonstrated \cite{BDJ75,KDE97}.

Electrically tunable polarization optical filters based on the Lyot filter scheme, using the birefringent effect of electrically controlled
liquid crystalline layers instead of solid retarders have been investigated \cite{Shabtay01, Abdulhalim01, Abdulhalim02, Abdulhalim03, Abdulhalim04, Sharp01} and are already commercially available. Due to the use of polarizers and the limited transmission of liquid crystals, however, the overall peak transmission of filters using higher number of polarizers and/or liquid crystal elements is relatively poor.

Stacked composite plates are mathematically equivalent to composite pulses in quantum physics, which are widely used in nuclear magnetic resonance (NMR) \cite{L86} and since recently, in quantum optics \cite{Haffner,Timoney}.
This similarity stems from the formal analogy between the Schr\"odinger equation for a two-state quantum system and the equation for the evolution of the Jones polarization vector \cite{J41}.
It has been used by Ardavan \cite{Ardavan} who proposed to use the BB and NB composite pulses of Wimperis \cite{Wimperis} to design BB and NB WPs. In a recent work \cite{IRV12}, the analogy between the polarization Jones vector and the quantum state vector was used to propose a calculation method to derive the twisting angles of a stack of retarders with arbitrarily accurate BB polarization retarders, which promise to deliver very high polarization conversion fidelity in an arbitrarily broad range of wavelengths.

In this paper, we follow the theoretical proposal of Ivanov \emph{et al.} \cite{IRV12} to design and demonstrate experimentally stacked composite plates, which, when rotated at specific angles, act either as BB half-wave retarders or tunable NB filters.
In contrast to the earlier demonstration by Peters \emph{et al.}~\cite{Peters}, here we use the more flexible single-pass setup rather than the double-pass setup used there.
Moreover, we use multiple-order WPs, which are cheaper and easier to produce, and are therefore advantageous for practical reasons.
We also use a lamp as a light source, rather than lasers as used in Ref.~\cite{Peters}, which allows us to obtain more detailed transmission profiles in a wider wavelength range.
Furthermore, the developed model shows how to \emph{tune} the NB filter to \emph{any} desired wavelength (other than the design wavelength $\lambda_0$) by merely selecting suitable rotation angles.

In Sec.~\ref{sec:theory}, we present the theoretical method used to design our polarization BB retarders and NB filters.
The experimental apparatus is described in details in Sec.~\ref{sec:experiment}.
In Sec.~\ref{sec:results}, we show the measured transmittance spectra of our composite filters and BB retarders.
In this section, we also show how to tune the central wavelength of the NB filters by simply selecting suitable rotation angles.
In Sec.~\ref{sec:conclusions}, we present the conclusions.

\section{Theory and numerical calculations}\label{sec:theory}

Here we will summarize the basic theory of composite optical retarders.
In the Jones calculus \cite{J41}, a single birefringent retarder is described by the matrix
\begin{equation}
\textbf{J}_{\alpha}(\varphi) =
R(-\alpha)
\left[
\begin{array}{cc}
e^{i\varphi/2} & 0 \\
0 & e^{-i\varphi/2}%
\end{array}%
\right]
R(\alpha),
\end{equation}
where
\begin{equation}
R(\alpha)=\left[
\begin{array}{cc}
\cos \alpha & \sin \alpha \\
-\sin \alpha & \cos \alpha%
\end{array}%
\right].
\end{equation}
We use the linear polarization basis, a pair of orthogonal polarization vectors, where $\alpha$ is the rotation angle of the retarder's optical axis
and $\varphi=\varphi(\lambda)$ is the retardation accumulated between the ordinary and the extraordinary rays passing through the retarder.
A stack of $N$ retarders is described by the composite Jones matrix
\begin{equation}
\textbf{J}^{\left( N\right) }(\retardation)=\textbf{J}_{\alpha _{N}}\left( \varphi _{N}\right) \textbf{J}_{\alpha_{N-1}}\left( \varphi _{N-1}\right) \cdots \textbf{J}_{\alpha _{1}}\left( \varphi_{1}\right),
\label{overall Jones matrix}
\end{equation}
where the light passes through the plates described by $\textbf{J}_{\alpha _{j}}\left( \varphi _{j}\right)$ in the order of ascending index $j$, i.e. from right to left,
 and $\retardation$ is the overall composite retardation.
For the sake of brevity, we denote $\textbf{J}^{\left( N\right) }(\retardation)=\jones$.
For practical purposes we take all plates to be QWP (Q) with respect to the design wavelength $\lambda_0$.
Thus we set $\varphi_j=\varphi_0$, where $\varphi_0=\pi/2$ for $\lambda_0$.
This allows us to use commercially available standard WPs.

\begin{figure}[ht]
\centering
\includegraphics[width=0.9\columnwidth]{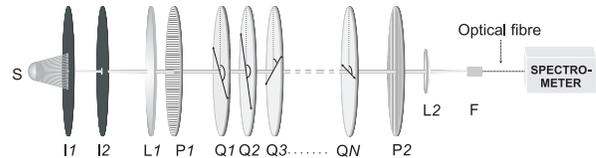}
\caption{Experimental setup. The source $S$, irises $I_{1}$ and $I_{2}$, lens $L_{1}$ and polarizer $P_{1}$ form a collimated beam of white polarized light. Polarizer $P_{2}$ and lens $L_{2}$ focus the  beam of output light onto the entrance $F$ of an optical fibre connected to a spectrometer.
The composite retarders and filters are constructed by a stack of multiple-order QWPs $(Q_{n})$.}
\label{fig:fig1}
\end{figure}

We thereby assume the following configuration of $N$ plates each rotated at an angle $\alpha_j$ (cf.~Fig.~\ref{fig:fig1}),
\begin{equation}
\label{sequence}
\text{Q}_{\alpha_{1}}\text{Q}_{\alpha_{2}} \cdots \text{Q}_{\alpha_{n-1}} \text{Q}_{\alpha_{N}}.
\end{equation}
The overall composite retardation is $\retardation=2\cos^{-1}\text{Re\,}J_{11}$. The element $J_{11}$ shows the fraction of light intensity which survives when light of particular polarization is passed through the filter.
The retardation profile is symmetric relative to the target retardation $\retardation=\pi$, i.e., $\retardation(\varphi)=\retardation(2\pi-\varphi)$.

By varying the $N$ rotation angles $\alpha_j$ we obtain different half-wave composite BB retarders and NB filters.

For \textit{BB retarders} the angles $\alpha_j$ satisfy
\begin{equation}
\text{max}(\left|\retardation(\varphi)/\pi-1\right|^2)\leq \eps, \quad \varphi\in \left[\varphi_\text{min},\pi-\varphi_\text{min}\right].
\label{eqn:numeric1}
\end{equation}

This
 guarantees that there is a range of single plate retardations between $\varphi_\text{min}$ and $\pi-\varphi_\text{min}$ (with $\varphi=\pi/2$ corresponding to $\lambda_0$), where the composite retardation $\retardation$ remains close to $\pi$.
We determine numerically those angles $\alpha_j$, which minimize $\varphi_\text{min}$, giving the broadest possible range of high-quality retardation.

On the contrary, a \textit{NB filter} must transmit only a small spectral region around $\lambda_0$.
This imposes the following relation on the angles $\alpha_j$:
\begin{equation}
\text{max}(\left|J_{12}\right|^2) \leq \eps, \quad  \varphi\in \left[0, \varphi_{\max} \right]\cup \left[\pi-\varphi_{\max}, \pi\right],
\label{eqn:numeric2}
\end{equation}
where $\left|J_{12}\right|^2$ represents the filter's intensity transmittance from horizontal to vertical polarization, as known from Jones calculus. \black
We seek those angles $\alpha_j$, which maximize $\varphi_{\max}$, thereby giving the narrowest possible transmission window around $\pi/2$.
Outside this window the transmission is maintained below $\eps$.
For both the BB and NB retarders we choose $\eps=1\%$.

We set the optical axes of the composite retarders and filters at angle 0$^\circ$ by setting $\left|J_{12}\right|=1$ at $\varphi=\pi/2$.
This imposes an additional constraint on the angles $\alpha_j$ (cf.~Fig.~\ref{fig:fig1}), thereby leaving $N-1$ angles $\alpha_j$ free to vary.

The more angles we can vary, the easier it is to fulfill Eqs. \eqref{eqn:numeric1} and \eqref{eqn:numeric2}. Therefore longer sequences \eqref{sequence} of larger number $N$ of constituent wave plates provide larger bandwidths of the broadband retarders or smaller bandwidths for the narrowband filters.

It is remarkable that we can \emph{tune} the wavelength of the filter, i.e. we can center the transmission window at a wavelength $\lambda^\prime$ other than the design wavelength of the single plate $\lambda_0$, just by varying the values of the rotation angles $\alpha_i$.
The latter are obtained from Eq.~\eqref{eqn:numeric2}, where now an offset in the argument of $J_{12}$ is introduced: $J_{12}(\varphi)\rightarrow J_{12}(\varphi^\prime)$.
Here $\varphi^\prime$ corresponds to the desired new wavelength $\lambda^\prime$: $\varphi^\prime = 2\pi L(n_e-n_o)/\lambda^\prime$.

For the numerical optimization we use Newton's gradient-based method.
Numerous solutions to Eqs. \eqref{eqn:numeric1} or \eqref{eqn:numeric2} exist depending on the values of $\varphi_{\min}$ or $\varphi_{\max}$.
To determine the optimal solutions we gradually decrease $\varphi_{\min}$ or increase $\varphi_{\max}$, until we reach the least $\varphi_{\min}$ or the largest $\varphi_{\max}$, which still satisfy Eqs. \eqref{eqn:numeric1} or \eqref{eqn:numeric2}. Because we use a local optimization algorithm, we iteratively pick the initial parameter values using a Monte-Carlo scheme.
Although the calculations are made for zero-order plates ($m$=0), they are valid for multi-order plates too as the matrix $J_\alpha(\varphi)$ and the ensuing matrix $J^{(N)}(\Phi)$ are periodical in the retardation $\varphi$.

The calculated results for the rotation angles $\alpha_j$ of the optical axes of the individual plates for several composite sequences are given in the captions of the respective figures below.


\section{Experiment}\label{sec:experiment}

We have tested experimentally the calculated BB retarders and NB filters by using the setup shown in Fig.~\ref{fig:fig1}.
A 10 W Halogen-Bellaphot (Osram) lamp with DC power supply was used as a source of white light.
A collimated light beam was made by using a variable iris $I_{1}$ with an aperture less than 0.5 mm positioned at the focus of a plano-convex lens $L_{1}$ ($f$=150 mm).
A second iris diaphragm $I_{2}$ was used to avoid outer rings of the beam pattern.
The diameter of the beam, measured at a distance of 2 m, was about 2 mm.
The light beam was polarized in the horizontal plane by a linear polarizer $P_{1}$.
A second polarizer $P_2$ was placed at a distance of 0.7 m and served as an analyzer during the experiments.
Both polarizers (Glan-Tayler, 210-1100 nm) were borrowed from a Lambda-950 spectrometer.
A second plano-convex lens $L_{2}$ ($f$=20 mm) and a two-axis micro-positioner were used to focus the light beam onto the optical fibre
entrance $F$ connected to a grating monochromator (Model AvaSpec-2048 with controlling software AvaSoft 7.5).
Multi-order QWPs (WPMQ10M-780, Thorlabs) were used to build the composite retarders as described above.
At wavelength 780 nm the multi-order QWPs are designed to be 11.25 waves, while at 765 nm they work as HWPs.
Each WP (aperture of $1^{\prime\prime}$) was assembled onto a separate RSP1 rotation mount.
The optical axes of all the WPs were determined with an accuracy of $1^{\circ}$.
With the used light source and monochromator, we could obtain reliable spectral transmission data in the range of 400-830 nm.

\begin{figure}[t]
\centering
\includegraphics[width=0.9\columnwidth]{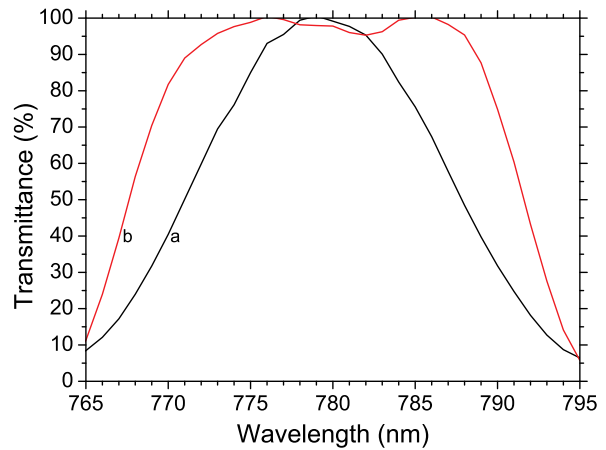}
\caption{Measured transmittance for BB composite HWP formed of a sequence of QWPs:
 a) reference spectrum of a HWP made of two QWPs;
 b) BB HWP retarder made of 6 QWPs with rotation angles (45.5$^\circ$, 78.5$^\circ$, 76.7$^\circ$, 15.5$^\circ$, 17.7$^\circ$, 45.4$^\circ$).
}
 \label{fig:fig2}
\end{figure}

The composite retarders were assembled as a set of multi-order WPs, each of them rotated at specific angles provided by the calculations, as described above.
The WPs were slightly tilted \cite{Peters} to reduce unwanted reflections.
We used a single beam spectrometer, thus all experiments started with measurement of the dark and reference spectra. The dark spectrum, taken with the light path blocked is further automatically used to correct for hardware offsets. The reference spectrum is usually taken with the light source on and a blank sample instead of the sample under test. In our case, however, we have measured the transmission spectrum of the already assembled composite retarder, but  the axes of the polarizer $P_{1}$, the analyzer $P_{2}$ and the fast axis of the single WPs were all set parallel. The measured light spectrum was used as a reference for the subsequent measurements with the waveplates rotated at their respective theoretically calculated angles $\alpha_j$, and the analyzer $P_{2}$ set to $90^{\circ} $. The transmission spectrum of the so-obtained composite retarder was recorded and scaled to the reference spectrum. The unavoidable losses due to reflections and absorptions from any single waveplate were thereby taken into account. The obtained spectrum  predominantly depends on the effective retardance of the composite retarder.

When the composite half-wave retarder was built by QWPs, the measured transmission spectra for both BB and NB retarders were compared
with the spectrum of two QWPs assembled to act as a single HWP.
In the cases when the WPs WPMQ10M-780 were used as HWPs at 765 nm, the measured transmission spectrum of a single WP was used to demonstrate the broadening or narrowing effect.

According to the theoretical model there are many suitable sets of angles for the composite retarders.
In all cases we have tested experimentally up to 10 different sets of angles for each retarder type, but here, for the sake of brevity we present only the representative results.

\begin{figure}
\includegraphics[width=0.9\columnwidth]{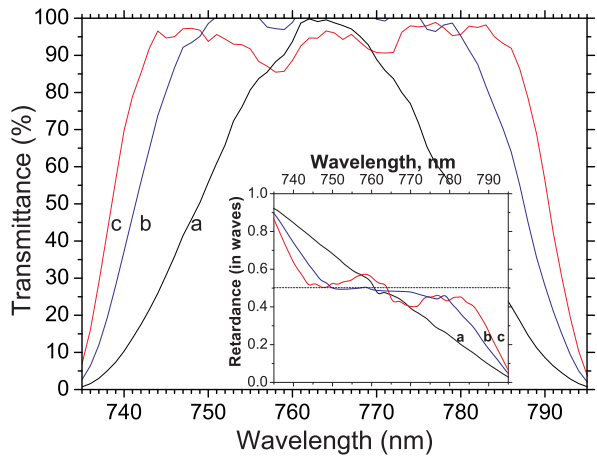}
\caption{Measured transmittance for BB composite HWP assembled by a sequence of QWPs used as HWPs at 765 nm.
 a) reference spectrum of one QWP;
 b) BB HWP retarder made of 3 QWPs with rotation angles (14.7$^\circ$, 164.4$^\circ$, 14.7$^\circ$);
 c) BB HWP retarder made of 5 QWPs with rotation angles (7.5$^\circ$, 172.5$^\circ$, 14.2$^\circ$, 172.9$^\circ$, 8.6$^\circ$).
The inset presents the retardance.}
\label{fig:Fig201}
\end{figure}

\section{Experimental results}\label{sec:results}

\subsection{Broadband polarization retarders}

In this section we will demonstrate that using the calculation method described above composite half-wave retarders with enhanced wavelength range can be assembled using a set of multi-order WPs. The design wavelength of the used multi-order (11.25 waves) QWPs (WPMQ10M-780) is 780 nm.
These WPs could also be used at 765 nm as HWPs (cf. the data sheet of WPMQ10M-780, Thorlabs).
We will demonstrate here how these commercial QWPs can be used to construct a composite BB HWP for both the designed wavelength 780 nm and 765 nm.

First, a composite HWP was made using six multi-order QWPs WPMQ10M-780.
The QWPs were twisted to the respective angles listed in the caption of Fig.~\ref{fig:fig2}.
Figure~\ref{fig:fig2} shows the transmittance curve.
For comparison, the transmittance spectrum of a HWP build up of two QWPs with parallel axes is shown too.
The significant wavelength range broadening yielded with the composite HWP retarder is evident: the width of the 90\%-efficiency range is nearly tripled.

In a second experiment, a BB HWP for 765 nm was composed using the same QWPs as above, but now used as HWPs at 765 nm.
Representative transmittance spectra of two composite BB HWPs, assembled with 3 and 5 HWPs, respectively, are presented in Fig.~\ref{fig:Fig201}.
The transmittance spectrum of a single HWP at 765 nm is shown for comparison. The calculated retardance is shown as inset in Fig.~\ref{fig:Fig201}.
As in Fig.~\ref{fig:fig2}, we find significant wavelength range broadening.
As expected, the composite HWP with higher number of constituent WPs has a wider wavelength range.

\begin{figure}
\centering
\includegraphics[width=0.9\columnwidth]{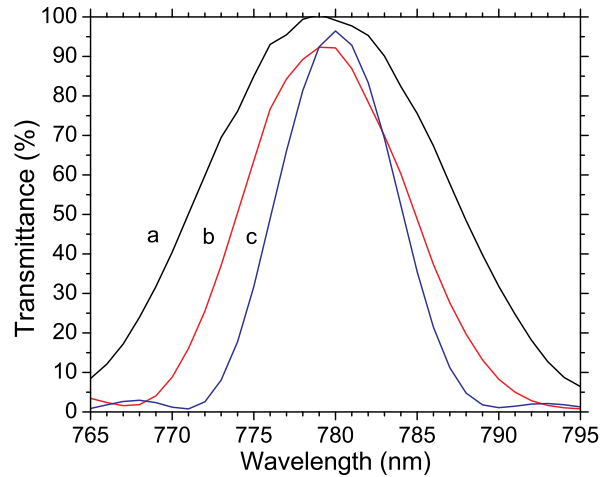}
\caption{Measured transmittance for  NB composite filters.
 a) reference spectrum of two QWPs;
 b) filter of 5 QWPs, rotation angles (43.7$^\circ$, 176.9$^\circ$, 170.1$^\circ$, 119.3$^\circ$, 80.2$^\circ$);
 c) filter of 6 QWPs, rotation angles (165.3$^\circ$, 167$^\circ$, 19.7$^\circ$, 18.6$^\circ$, 166.4$^\circ$, 166.1$^\circ$).
}
\label{fig:fig3}
\end{figure}

\subsection{Composite polarization filters}

Standard multi-order WPs are NB retarders.
Using a proper set of rotation angles, a composite NB retarder can be assembled with much narrower wavelength range.
Here we demonstrate how with the same set of QWPs, as the one used above, but rotated at different angles, a NB HWP can be assembled.
Figure~\ref{fig:fig3} shows representative transmittance curves for composite NB filters constructed of 5 and 6 QWPs, respectively.
The 6-plate composite NB filter reduces the width (at half-maximum) of the transmittance spectrum by a factor of almost 3.
The comparison of the transmittance spectra (curves \emph{b} and \emph{c}) shows that the filter bandwidth is narrower for a higher number of WPs.
Even with this limited number of WPs one can produce filters with a bandwidth of about 7 nm.
There is no fundamental physical limit (except the divergence of the light beam), which can prevent us from going to as narrow width of the transmission profile as we like. With sufficiently many WPs, one can reduce the bandwidth even below 1 nm for sufficiently collimated light beam.

\begin{figure}
\includegraphics[width=0.9\columnwidth]{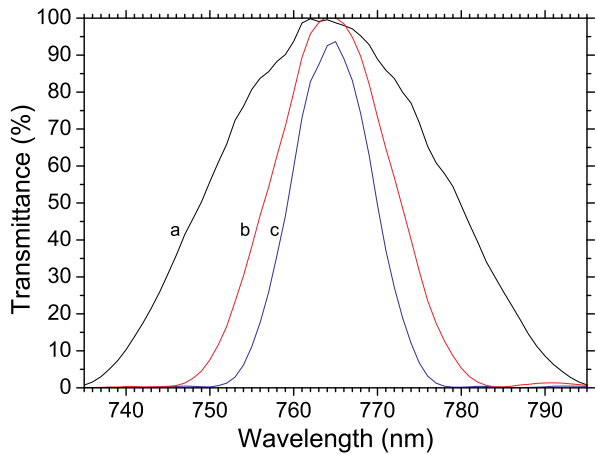}
\caption{Measured transmittance for composite filters.
 a) reference spectrum of one HWPs;
 b) filter of 3 HWPs, rotation angles (14.7$^\circ$,~45.1$^\circ$,~75.3$^\circ$);
 c) filter of 5 HWPs, rotation angles (37.7$^\circ$,~83.1$^\circ$,~49.9$^\circ$,~1.9$^\circ$,~38.2$^\circ$).
}
\label{fig:Fig202}
\end{figure}

As for BB retarders, we have calculated sets of angles for the NB HWPs assembled using the QWPs as HWPs at 765 nm.
The results are shown in Fig.~\ref{fig:Fig202}.
Again, a reduction of the transmittance bandwidth by a factor of 3 is established with the 6-plate composite filter.

\subsection{Tunable composite polarization filters}

\begin{figure}[t]
\includegraphics[width=0.9\columnwidth]{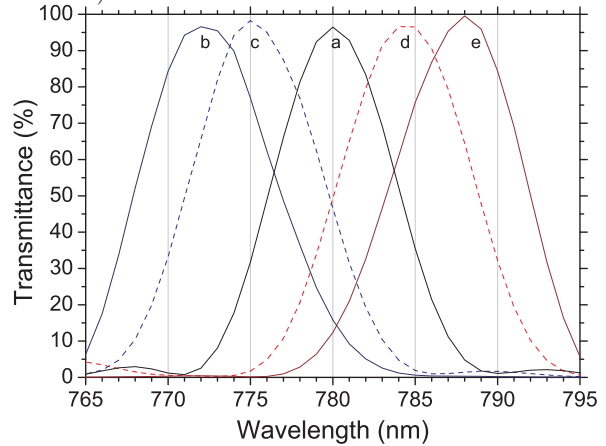}
\caption{Measured transmittance of composite filters made of the same set of 6 QWPs but for different rotation angles.
The central wavelength is tuned to:
 a) 780 nm (design wavelength, no tuning, retardation $\varphi^\prime=0.50\pi$);
 b) 772 nm ($\varphi^\prime=0.25\pi$);
 c) 775 nm ($\varphi^\prime=0.35\pi$);
 d) 784 nm ($\varphi^\prime=0.65\pi$);
 e) 788 nm ($\varphi^\prime=0.75\pi$).
The rotation angles are:
a) (165.3$^{\circ}$, 167.1$^{\circ}$, 19.7$^{\circ}$, 18.6$^{\circ}$, 166.4$^{\circ}$, 166.1$^{\circ}$);
b) (8.9$^{\circ}$, 158.1$^{\circ}$, 158.9$^{\circ}$, 116.1$^{\circ}$, 51.5$^{\circ}$, 34.5$^{\circ}$);
c) (26.4$^{\circ}$, 154.7$^{\circ}$, 178.6$^{\circ}$, 1.3$^{\circ}$, 20.2$^{\circ}$, 158.8$^{\circ}$);
d) (60.4$^{\circ}$, 14.1$^{\circ}$, 175.7$^{\circ}$, 178.4$^{\circ}$, 154.2$^{\circ}$, 110.6$^{\circ}$);
e) (6.2$^{\circ}$, 25.4$^{\circ}$, 50.2$^{\circ}$, 56.1$^{\circ}$, 39.0$^{\circ}$, 13.4$^{\circ}$).
 }
\label{fig:fig4}
\end{figure}

As already shown, very narrow-band optical filters can be assembled using a stack of QWPs or HWPs with properly chosen rotation angles.
Furthermore, the calculation method presented above can be also applied to solve the problem of making these filters tunable.
The tunability was tested for a composite HWP filter composed of 6 QWPs.
The measured set of transmittance spectra is shown in Fig.~\ref{fig:fig4}.
It is seen that by choosing properly the rotation angles the filter's spectral band can be moved at will.
Remarkably, the bandwidth of the transmittance curves does not change with the tuning of the filter.
Even more, one could use the multi-order WPs as QWPs or as HWPs at different wavelengths as shown in the previous section, thus constructing tunable narrow-band filters in a wide wavelength range.

\section{Conclusion}\label{sec:conclusions}

We have demonstrated that by using a stack of ordinary (chromatic) multi-order WPs, which act as HWP at 765 nm and QWP at 780 nm, one can build efficient polarization BB (achromatic) half-wave composite retarders and tunable NB polarization filters.
By rotating the same set of WPs at specific angles we have constructed BB HWP around 780 nm, BB HWP around 765 nm, and tunable HWP filter at several wavelengths from 765 nm to 788 nm.
Better (broader) BB retarders can be built with lower-order WPs.
Polarization filters with narrower bandwidth (FWHM), even below 1 nm, can be constructed with sufficiently many high-order (the higher the better) WPs.

\acknowledgements

This work was supported by the Bulgarian NSF Grant DRila - 01/4 and the European Community's Seventh Framework Programme under grant agreement No. 270843 (iQIT).


\end{document}